\documentclass[twocolumn]{aastex631}

\usepackage{graphicx,comment}

\newcommand{\HST}{\emph{HST}}
\newcommand{\JWST}{\emph{JWST}}

\newcommand{\Muv}{\ensuremath{M_\mathrm{UV}^{ }}}

\definecolor{red}{RGB}{225,50,50}

\begin{document}

\title{An Extremely-red, UV-bright, and Extended Galaxy at {$z\sim6$} in PRIMER/UDS: \\ An Early Massive Galaxy Caught Quenching after an Obscured Starburst?}

\author{Nadara Hudson}
\affiliation{Department of Astronomy, University of Virginia, Charlottesville, VA 22903, USA}

\author{Ryan Endsley}
\affiliation{Department of Astronomy, University of Texas at Austin
Austin, TX 78712, USA}

\author{John Chisholm}
\affiliation{Department of Astronomy, University of Texas at Austin
Austin, TX 78712, USA}

\begin{abstract} 
JWST continues to reveal an astonishing number of massive quiescent galaxies at $z>4$, with number densities $\gtrsim10\times$ higher than model predictions. NIRSpec spectra imply that many of these systems underwent intense starburst episodes (SFR$\,\gtrsim300M_\odot$/yr), though direct evidence of such starbursts in the Gyr largely comes from exceptionally rare dusty star-forming galaxies (DSFGs) selected in the far-infrared. Here, we report the discovery of an extremely red ($\beta=-0.6$) yet UV-bright (F115W = 26.0 mag) $z\sim6$ star-forming system selected as a Lyman-break galaxy (LBG) over $\approx$500 arcmin$^2$ of deep NIRCam imaging. This galaxy (UDS\_43065) shows photometric colors implying a prominent Balmer break and strong H$\alpha$ emission, consistent with a dramatic burst of star formation (SFR$\,\approx\,500-1000\,M_\odot$/yr) occurring 5--10 Myr ago that formed 20--40\% of its total stellar mass ($\approx1.5\times10^{10}M_\odot$) with little activity since. This galaxy is one of only two objects with $M_\ast>10^{10}M_\odot$ across our full sample of 813 $z\sim6$ star-forming LBGs, and the only galaxy with a confident extremely-red UV slope ($\beta>-1$). UDS\_43065 is clearly resolved yet compact in F444W ($r_e=400\pm10$ pc) indicating a very high stellar mass surface density of log$(\Sigma_\mathrm{eff}/(M_\odot\,\mathrm{kpc}^{-2}))=10.25\pm0.13$ comparable to quenched $z\sim2-7$ galaxies. If the inferred star formation history (SFH) of UDS\_43065 is corroborated with further observations, this object would seemingly represent a rarely-seen transitional phase between massive DSFGs and passive systems in the first Gyr, helping resolve the puzzling abundance of early massive quenched galaxies. 

\end{abstract}

\section{Introduction} \label{sec:intro}
\JWST{} Near-infrared Camera (NIRCam; \citealt{Rieke2023}) imaging surveys have revealed several massive (\(M_\ast\sim10^{10-11}\,M_\odot\)) quiescent galaxy candidates within the first 1.5 billion years of cosmic history ($z>4$; e.g., \citealt{Carnall2023a,Valentino2023,Long2024,Baker2025b}), with empirical number densities often exceeding model predictions by over an order of magnitude \citep[e.g.,][]{Lagos2025}.
Spectroscopic follow-up with the Near-infrared Spectrograph (NIRSpec; \citealt{Jakobsen2022}) is continuing to confirm the nature of many photometric candidates \citep[e.g.,][]{Carnall2023b,Carnall2024,Kokorev2024b,deGraaff2025,Weibel2025}, solidifying the marked tension between models and observations \citep[e.g.,][]{Baker2025a}.
Resolving this discrepancy will require an improved understanding of the physical mechanisms driving the rapid assembly and quenching of massive galaxies.

New insights are emerging from the spectral energy distributions (SEDs) of the earliest massive quiescent galaxies ($z\sim5-7$) targeted with NIRSpec. 
In the context of modern stellar population synthesis (SPS) models, the spectra imply that much of their stellar mass formed during an intense starburst phase with star formation rate (SFR) $\gtrsim$300 $M_\odot$/yr \citep[e.g.,][]{Carnall2023b,Carnall2024,deGraaff2025,Weibel2025}. 
The stellar feedback from such an intense starburst phase may help exhaust the gas fuel supply for subsequent star formation, but the feedback efficiency is sensitive to the unknown duration and instantaneous SFR of that starburst phase \citep{Lagos2025}.
Massive quenched $z\gtrsim5$ galaxies are often observed $\gtrsim$300 Myr after the most recent starburst phase, and spectra quickly loose short-timescale SFH information after a burst \citep[e.g.][]{French2018,Leja2019}.
Consequently, it is very challenging to determine if the starburst phase in these early massive quiescent galaxies lasted of order $\sim$10 Myr or $\sim$100 Myr, resulting in order-of-magnitude uncertainties in the instantaneous SFR as well. 

A key goal over the coming years will be identifying a statistical sample of $z\gtrsim5$ massive galaxies that have very recently exited an intense starburst phase, where their SEDs encode the star formation history (SFH) surrounding and during the burst to much greater detail.
The primary challenge of this endeavor lies in the rarity of these galaxies.
Not only are early massive galaxies rare by nature of structure formation, but we aim to catch them within a short time interval after a starburst episode.
One path towards identifying these rare systems is to conduct targeted searches for bright, dusty galaxies at $z\gtrsim5$, since starbursts create huge amounts of dust that is subsequently destroyed or expelled \citep[e.g.,][]{Li2019}.
While dedicated far-infrared surveys have revealed several $z\gtrsim5$ dusty galaxies over the past decade, these objects are often extremely obscured in the UV and optical, thus precluding detailed SFH constraints \citep{Casey2014}.
For this reason, we have opted to search for dusty yet bright early galaxies using a Lyman-break selection, requiring that the selected systems remain luminous in the UV and optical.

In this initial work, we use an $\approx$500 arcmin$^2$ area with public \JWST{}, \textit{Hubble Space Telescope} (HST), and ground-based imaging to search for dusty, bright Lyman-break galaxies at $z\sim6$, and subsequently characterize their stellar masses and SFHs.
We report the discovery of a UV and optically bright (F115W = 26.0 AB mag; F444W = 23.9 AB mag), massive \((M_\ast\approx1.5\times10^{10}\,M_\odot)\), dusty (\(A_V\approx1.5\) mag; $\beta \approx -0.6$) Lyman-break galaxy at \(z\approx5.6\) that is clearly resolved in F444W.
We discuss how this galaxy, named UDS\_43065, fits into the broader context of reionization-era objects and how its inferred SFH may hold key clues to resolving the puzzling abundance of early massive quenched galaxies.

Throughout this work, all magnitudes are reported in the AB system \citep{Oke1983} and we adopt a \citet{Chabrier2003} stellar initial mass function (IMF) with limits of 0.1--300 $M_\odot$ as well as a flat $\Lambda$CDM cosmology with parameters $h=0.7$, $\Omega_\mathrm{M}=0.3$, and $\Omega_\mathrm{\Lambda}=0.7$.

\begin{figure*}[ht]
    \centering
    \includegraphics[width=0.95\textwidth]{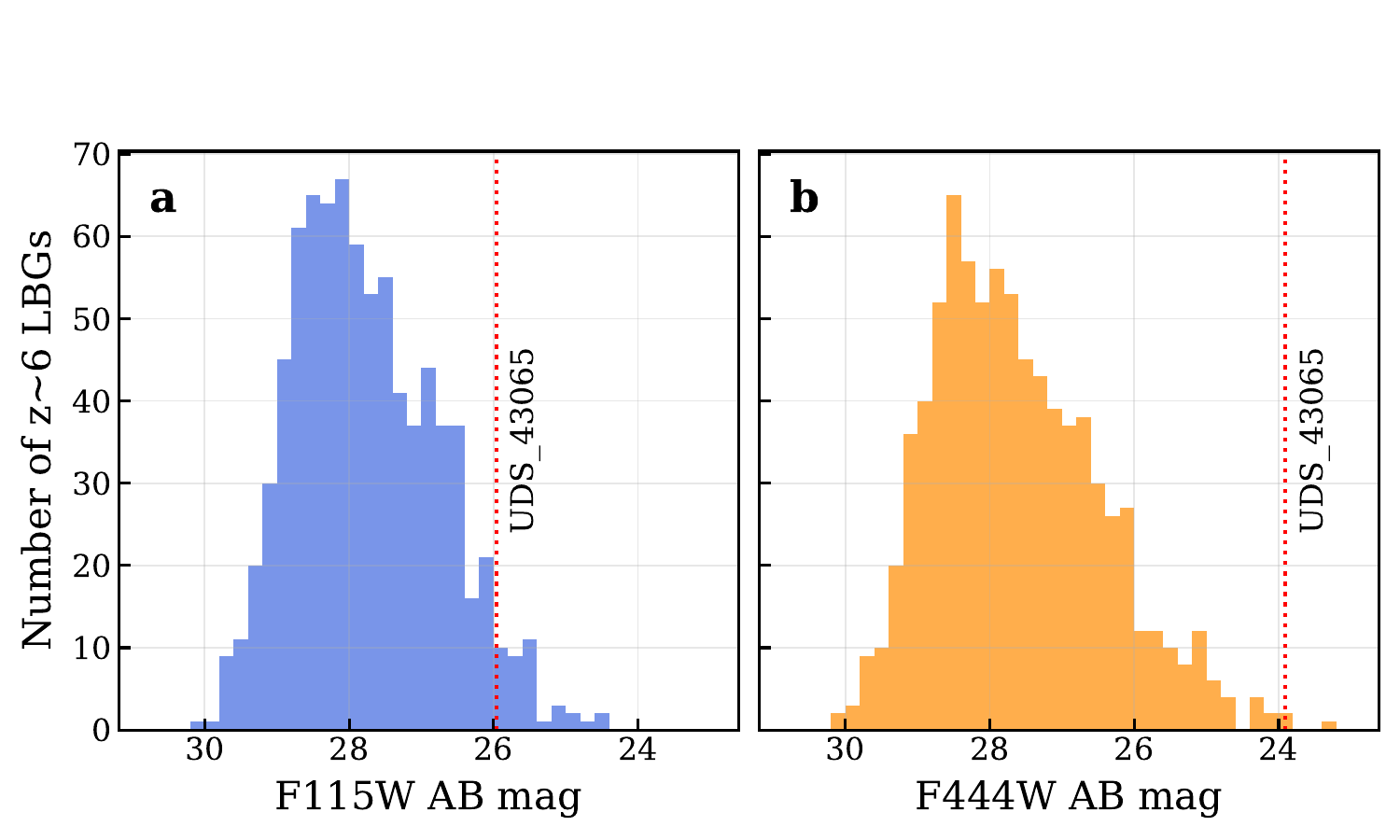} 
    \caption{Histograms of the F115W (left) and F444W (right) apparent magnitudes for our sample of 813 z$\sim$6 Lyman-break galaxies selected over $\approx$500 arcmin$^2$ of deep JWST/NIRCam imaging. The apparent magnitudes of UDS\_43065, the primary object discussed in this work, are shown with vertical dotted red lines.}
    \label{fig:magDistn}
\end{figure*}

\begin{figure*}[ht]
    \centering
    \includegraphics[width=1.0\textwidth]{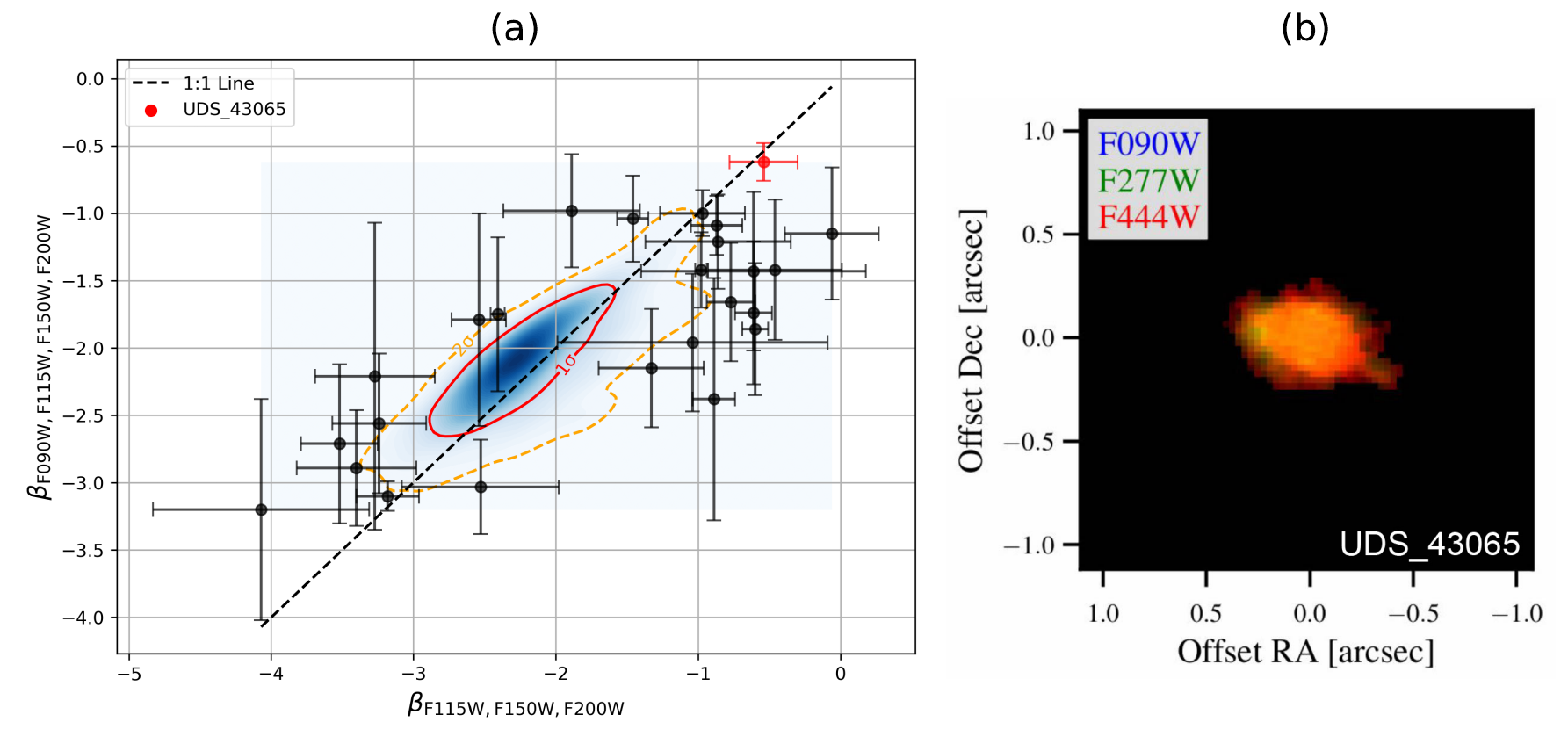}
    \caption
    {\textbf{(a)} A comparison of two rest-UV continuum slope ($F_\lambda \propto \lambda^\beta$) measurements among our $z\sim6$ LBG sample. The x-axis shows the UV slope measurement using the three NIRCam broadbands F115W, F150W, and F200W, while the y-axis shows the measurement obtained when also folding in F090W. We show the 1$\sigma$ and 2$\sigma$ contours capturing the bulk of most of our sample, while outliers beyond this range are represented as individual data points. Only one object in our sample (UDS\_43065) has an extremely red UV slope ($\beta > -1$) from both measurements. This galaxy is very bright in the UV (F115W = 26.0 mag) and hence has precise UV slope measurements firmly placing it in this extremely red regime. \textbf{(b)} An RGB color image of UDS\_43065 illustrating its very red nature and resolved morphology ($r_e = 400\pm10 \, \text{pc}$).}
    \label{fig:red_galaxy}
\end{figure*} 

\section{Sample Selection and SED Modeling}\label{sec:method}

In this initial study, we aim to begin identifying candidate dusty, massive Lyman-break galaxies in the first billion years ($z\geq5.5$) and subsequently characterize their stellar masses and recent SFHs. 
Given the rarity of such objects, we consider all five extragalactic blank fields that were part of the \HST{} Cosmic Assembly Near-infrared Deep Extragalactic Legacy Survey (CANDELS; \citealt{Grogin2011,Koekemoer2011}) and have now been observed with \JWST{}/NIRCam \citep[e.g.,][]{Casey2023,Eisenstein2023,Oesch2023,Donnan2024,Finkelstein2025}, delivering deep multi-band imaging spanning $\approx$0.4--5$\mu$m.
Because we aim to characterize the SFHs of our target galaxies, we restrict our Lyman-break selection to $z\sim6$ where the H$\alpha$ line (highly sensitive to SFR over the past 3--10 Myr) is probed by NIRCam imaging.

In the GOODS-S and GOODS-N fields, we use the $z\sim6$ ACS/F775W dropout selection criteria described in \citet{Endsley2024}. 
For the COSMOS and UDS fields (which lack deep, comprehensive ACS/F775W coverage), we supplement the \HST{} Advanced Camera for Surveys (ACS) and \JWST{}/NIRCam imaging data with deep ground-based optical imaging from the Subaru Hyper Surpime-Cam (HSC; \citealt{Aihara2022}) to identify robust $z\sim6$ Lyman-break galaxies. Specifically, we ensure that all $z\sim6$ dropouts in COSMOS and UDS show a dropout in ACS/F606W, ACS/F814W, HSC/nb816, and HSC/\textit{i} relative to the NIRCam/F090W band by enforcing the following color cuts:
\begin{itemize}
    \item F606W $-$ F090W $>$ 2.0,
    \item F606W $-$ F090W $>$ (F090W $-$ F150W) $+$ 2.0,
    \item F814W $-$ F090W $>$ 0.5,
    \item F814W $-$ F090W $>$ (F090W $-$ F115W) $+$ 0.5,
    \item nb816 $-$ F090W $>$ 0.7,
    \item nb816 $-$ F090W $>$ (F090W $-$ F150W) $+$ 0.7,
    \item $i$ $-$ F090W $>$ 1.2, 
    \item $i$ $-$ F090W $>$ (F090W $-$ F150W) $+$ 1.2, and
    \item F090W $-$ F150W $<$ 1.0.

\end{itemize}
In cases where the object is not detected in the dropout band  (S/N$<$1), we set the flux to the 1$\sigma$ upper limit in that band.
For objects with \HST{}/ACS F435W coverage, we additionally enforce S/N(F435W)$<$2 and that F435W $-$ F090W $>$ 1.5.
For those lacking F435W coverage, we enforce S/N($g$)$<$2 and that $g$ $-$ F090W $>$ 2.0.
We also enforce that every LBG is detected at S/N$>$5 in at least one NIRCam band, as well as S/N$>$3 in at least three NIRCam bands.

We use all NIRCam imaging publicly available as of February 2024 in these four fields (GOODS-S, GOODS-N, COSMOS, and UDS) and process the raw data through to final mosaics following the procedure described in \citet{Endsley2024}, yielding $\approx$500 arcmin$^2$ of imaging area over which we can apply the above selection cuts.
We exclude the EGS field from our analysis because, at the time, comprehensive F090W imaging was not publicly available.
The \JWST{}/NIRCam and \HST{}/ACS photometry are computed in Kron apertures following \citet{Endsley2024} while the ground-based photometry is computed in 1.2\arcsec{} diameter apertures after running neighbor subtraction using the high-resolution F090W imaging as a morphological prior (similar to the IRAC deblending procedure described in \citealt{Endsley2021}). 

To characterize the physical properties of the $z\sim6$ LGBs, we use two photoionization SED fitting codes: BayEsian Analysis of GaLaxy sEds (\textsc{beagle}; \citealt{Chevallard2016}) and \textsc{prospector} \citep{Johnson2021}. Each code employs a Bayesian framework to infer various galaxy properties including redshift, stellar mass, SFH, and dust optical depth.
We use both \textsc{beagle} and \textsc{prospector} to demonstrate that our main conclusions are not driven by the specific SPS models, photoionization frameworks, or priors specific to one code.

For both \textsc{beagle} and \textsc{prospector}, we adopt a \citet{Calzetti2001} dust attenuation law, a \citet{Chabrier2003} IMF from 0.1--300 $M_\odot$, and a binned (i.e., `non-parametric') SFH prior. 
For \textsc{prospector} we use the `bursty continuity' SFH prior described in \citet{Tacchella2022}, while for \textsc{beagle} we adopt the binned SFH prior implementation described in \citet{Endsley2025}. 
For both codes, the first seven lookback time bins are fixed to 0--3 Myr, 3--5 Myr, 5--10 Myr, 10--20 Myr, 20--30 Myr, 30--50 Myr, and 50--100 Myr (as motivated in \citealt{Endsley2025}) and then we use three additional bins equally spaced in log lookback time between 100 Myr and the age of the Universe at the estimated photometric redshift.
We adopt a log-uniform prior on the ionization parameter ($-4 \leq \mathrm{log}\, U \leq -1$) in both codes.
For \textsc{beagle}, the stellar and interstellar medium (ISM) metallicities are equivalent and given a log-uniform prior in the range $-2.2 \leq Z/Z_\odot \leq -0.3$, though the effective gas-phase metallicity is altered by dust depletion using a fixed dust-to-metal mass ratio of $\xi_d = 0.3$ (see \citealt{Gutkin2016}).
In the \textsc{prospector} fits, we adopt a log-uniform prior for the gas-phase metallicity in the range $-2.2 \leq Z/Z_\odot \leq -0.3$, while the stellar metallicity is set to be lower by a factor of 3--5 (uniform prior) given evidence of strong $\alpha$-enhancement at early epochs \citep[e.g.,][]{Sanders2020,Cullen2021}. 

In our analysis, we ignored any extremely compact object where the photometry was very poorly fit (best-fitting $\chi^2 > 100$) by the star-forming only photoionization models used in both codes, as such systems have been shown to likely be a sub-population of high-redshift AGN whose galaxy properties remain unclear \citep[e.g.,][]{Endsley2023,Matthee2024,Wang2024,Kokorev2024a,Akins2025,Kocevski2025,Ma2025}.
Following the procedures described in \citet{Endsley2021,Endsley2024}, we also removed a handful of extremely compact sources showing colors consistent with brown dwarf solutions.
This selection procedure resulted in a final sample of 813 $z\sim6$ LBG candidates across GOODS-N, GOODS-S, COSMOS, and UDS.
The F115W and F444W magnitudes of galaxies in this sample span $m\sim30$ at the faint end to $m\sim24$ at the bright end (see Fig. \ref{fig:magDistn}).

\section{Results}\label{sec:results}

Within our sample of 813 $z\sim6$ LBG candidates, we identify a single object that confidently shows short-wavelength NIRCam colors implying a very red rest-UV slope ($\beta > -1$ where $F_\lambda \propto \lambda^\beta$; see Fig. \ref{fig:red_galaxy}a).
This galaxy, named UDS\_43065 (RA = 34.396334, Dec = $-$5.266858), has measured colors of F090W$-$F200W = $1.12 \pm 0.11$ and F115W$-$F200W = $0.77 \pm 0.09$.
Since the measured UV colors can have large uncertainties when fitting with a small number of bands, or if Ly$\alpha$ contaminates one of the bands (here F090W for $z\sim6$ LBGs), we computed $\beta$ using two sets of filters. 
UDS\_43065 is the only galaxy in our sample with a very red rest-UV slope measured from all four broadbands between (and including) F090W and F200W ($\beta = -0.62 \pm 0.14$) as well as the three broadbands between F115W and F200W ($\beta = -0.54 \pm 0.24$).
Even though UDS\_43065 exhibits an extremely red UV color, it is among the brightest 5\% of LBGs in F115W (26.0 mag; see Fig. \ref{fig:magDistn}) and is also extremely bright in F444W (23.9 mag).
Given the exceptional nature of this system, it is our sole focus in this article.

\begin{figure*}[ht]
    \centering
    \includegraphics[width=0.95\textwidth]{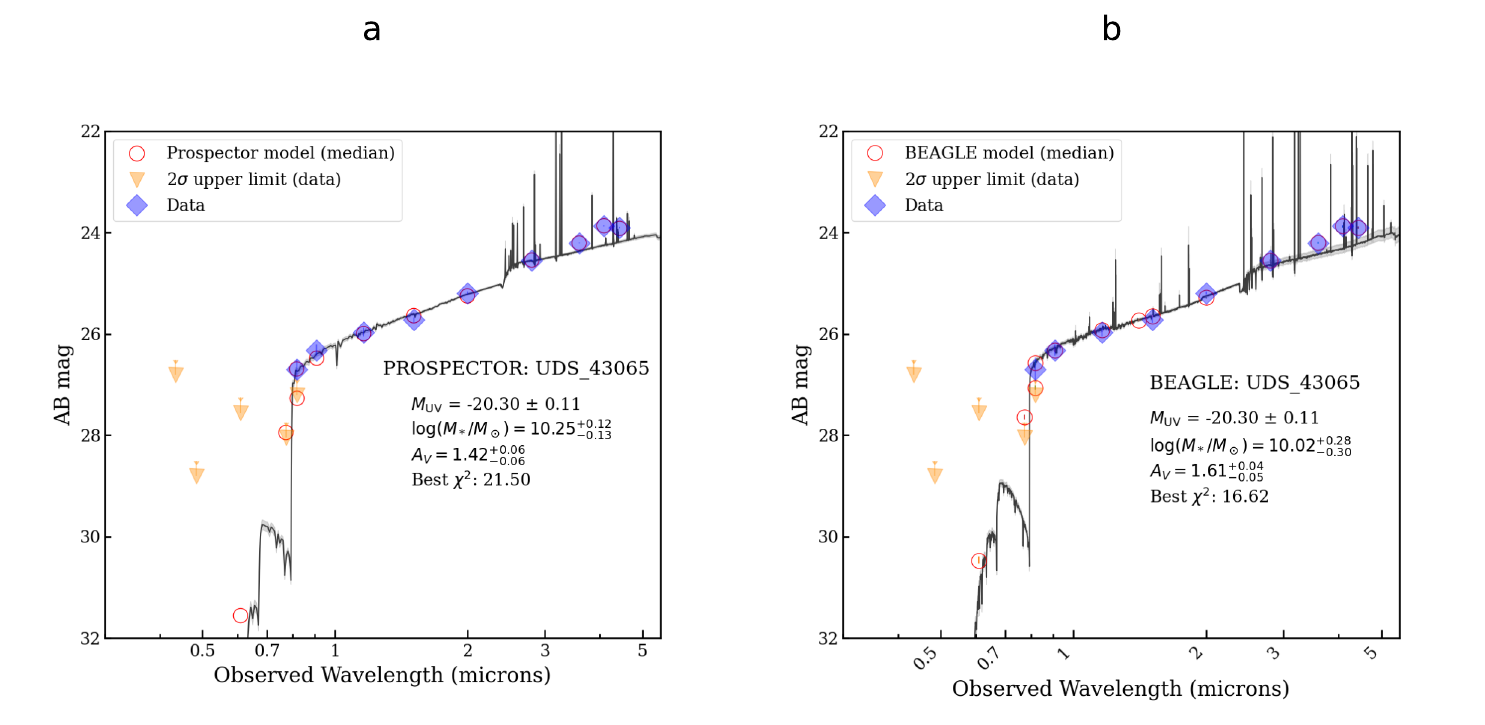} 
    \caption{\textsc{beagle} and \textsc{prospector} SED fits for UDS\_43065. Both codes are able to successfully reproduce the 14-band photometry, with best-fitting $\chi^2$ values of $\approx$20. The strong color between F200W and F277W is reproduced with a pronounced Balmer break ($F_{\nu,4200}/F_{\nu,3500} \approx 1.3-1.4$), implying that relatively old B and A-type stars are contributing substantially to the continuum light. Additionally, both \textsc{beagle} and \textsc{prospector} infer strong H-alpha emission (EW$\approx$400-500 \AA{}) in the F410M band, a hallmark of active star formation fueled by young, hot O-type stars. The codes consistently infer a high rest-frame V-band dust attenuation of  $A_V\approx1.5$ mag, as expected given the extremely red rest-UV color of this galaxy.}
    \label{fig:SEDs}
\end{figure*}

The extremely deep narrowband optical imaging from the HSC Subaru Strategic Program provides high confidence that UDS\_43065 lies at $z\sim6$.
UDS\_43065 is undetected (S/N$<$2) in the HSC/nb816 narrowband with a 2$\sigma$ upper limiting magnitude of $>$27.2 (and undetected in every band blueward of nb816), but is then 0.9 mags brighter in the immediately adjacent NIRCam/F090W band.
The red NIRCam colors of UDS\_43065 are insufficient to explain such a sharp spectral discontinuity around 0.8$\mu$m since the F090W$-$F115W color is only a moderate 0.35$\pm$0.14 mag.
As discussed in more detail below, UDS\_43065 shows NIRCam colors at 2--5$\mu$m that are consistent with a strong Balmer break and significant H$\alpha$ nebular emission at $z\approx5.6$, precisely the redshift where the Ly$\alpha$ break would fall at $\approx$0.8$\mu$m.
Consequently, both the \textsc{beagle} and \textsc{prospector} fits to the ACS+HSC+NIRCam photometry of UDS\_43065 confidently return a photometric redshift of $z_\mathrm{phot} \approx 5.6$.
At this redshift, the UV slope measurements are consistent with a rest-frame V-band attenuation of $\approx$1.4--1.6 mag in context of the \textsc{beagle} and \textsc{prospector} SED fits (see Fig. \ref{fig:SEDs}). We note that both \textsc{beagle} and \textsc{prospector} yield SED solutions that well match the measured photometry, with best-fitting $\chi^2$ values of 16.6--21.5 across the 14 fitted data points (Fig. \ref{fig:SEDs}).

UDS\_43065 shows a clearly resolved morphology in the NIRCam imaging (see Fig. \ref{fig:red_galaxy}b) which is well described by a 2D S{\'e}rsic model.
After fitting the F444W data, we recover a S{\'e}rsic index of 1.8$\pm$0.1 and a half-light radius of 0.66$\pm$0.02 arcsec, which translates to 400$\pm$10 pc at $z_\mathrm{phot}=5.55$.
These morphological parameters are broadly consistent with that of similarly UV-luminous (though often much bluer) $z\sim6$ LBGs derived from HST imaging \citep[e.g.,][]{Shibuya2015,CurtisLake2016}, though UDS\_43065 is $\approx$0.4 dex smaller than is typical at its \Muv{}.
UDS\_43065 shows no clear evidence of multiple clumps at the existing depth of the NIRCam imaging.

At the photometric redshift of $z\approx5.6$ implied by the HSC/nb816 dropout, the 3648 \AA{} Balmer break would lie at an observed wavelength of $\approx$2.4$\mu$m between the F200W and F277W bands.
The F277W flux density of UDS\_43065 (548$\pm$6 nJy) is significantly ($\approx$10$\sigma$) higher than that obtained by extrapolating the UV slope fits (475--490 nJy), consistent with a strong Balmer break. 
We also measure a significantly higher flux density in F410M than either F356W or F444W, consistent with strong H$\alpha$ emission boosting the brightness in F410M.
Indeed, both the \textsc{beagle} and \textsc{prospector} fits to the full HSC+ACS+NIRCam photometric dataset imply that UDS\_43065 possesses a strong Balmer break and significant H$\alpha$ emission (see Fig. \ref{fig:SEDs}).
Specifically, we infer a dust-corrected Balmer break strength of $F_{\nu,4200}/F_{\nu,3500} = 1.3 \pm 0.1$ from \textsc{beagle} and $F_{\nu,4200}/F_{\nu,3500} = 1.4 \pm 0.1$ from \textsc{prospector}.
The rest-frame H$\alpha$ EWs are inferred to be 490$\pm$150 \AA{} and 470$\pm$50 \AA{} from \textsc{beagle} and \textsc{prospector}, respectively.

\begin{figure}[ht]
    \centering
    \includegraphics[width=\columnwidth]{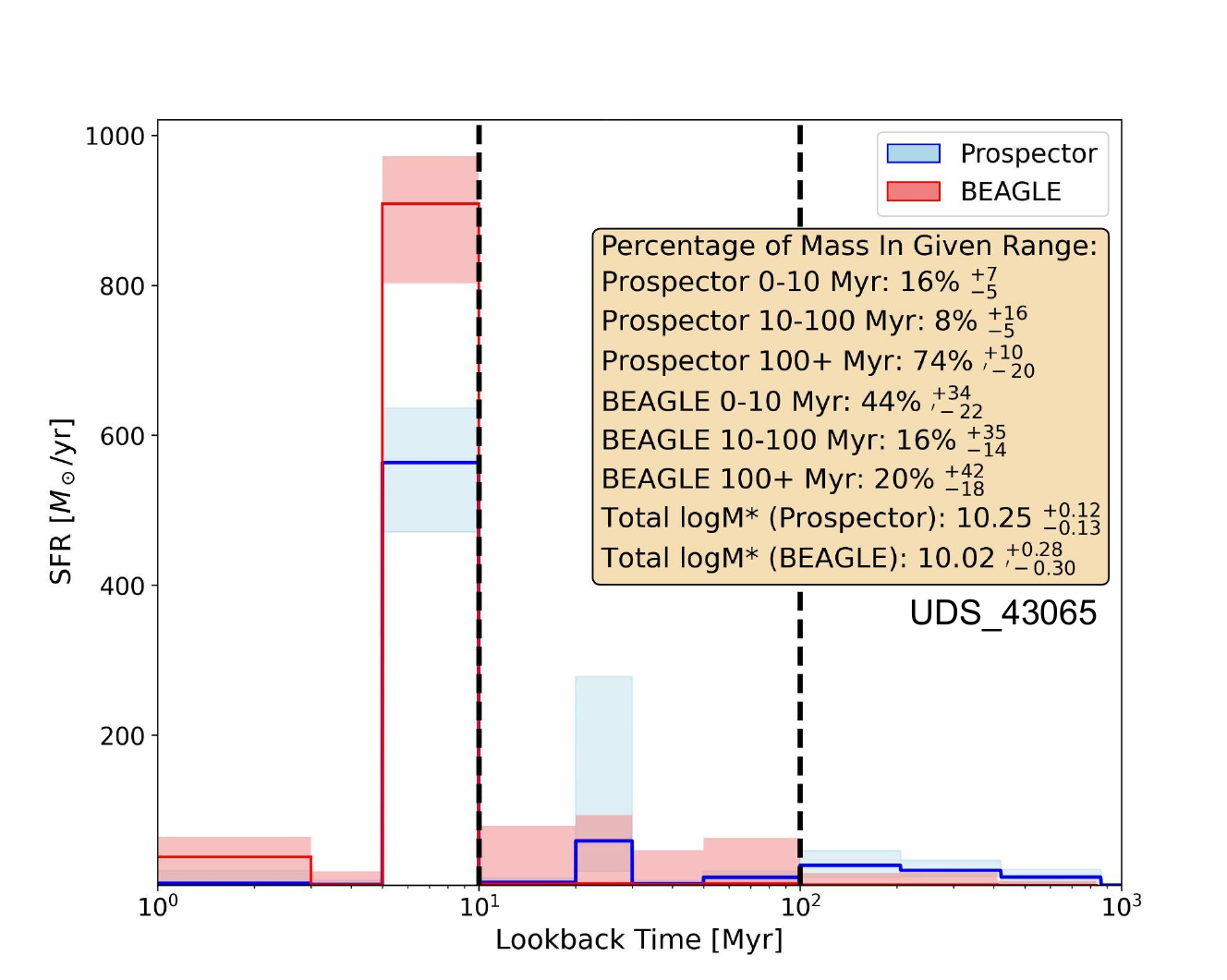}
    \caption{The inferred SFHs of UDS\_43065 from \textsc{prospector} (blue) and \textsc{beagle} (red). Both SED fits imply that UDS\_43065 experienced a dramatic burst of star formation $\approx$5--10 Myr ago, with a SFR $\approx500–1000\,M_\odot$/yr, and has been relatively inactive since. Approximately 20--40\% of the galaxy’s mass is inferred to have formed during this event.}
    
    \label{fig:SFH}
\end{figure}

\begin{figure*}[ht]
    \centering
    \includegraphics[width=0.70\textwidth]{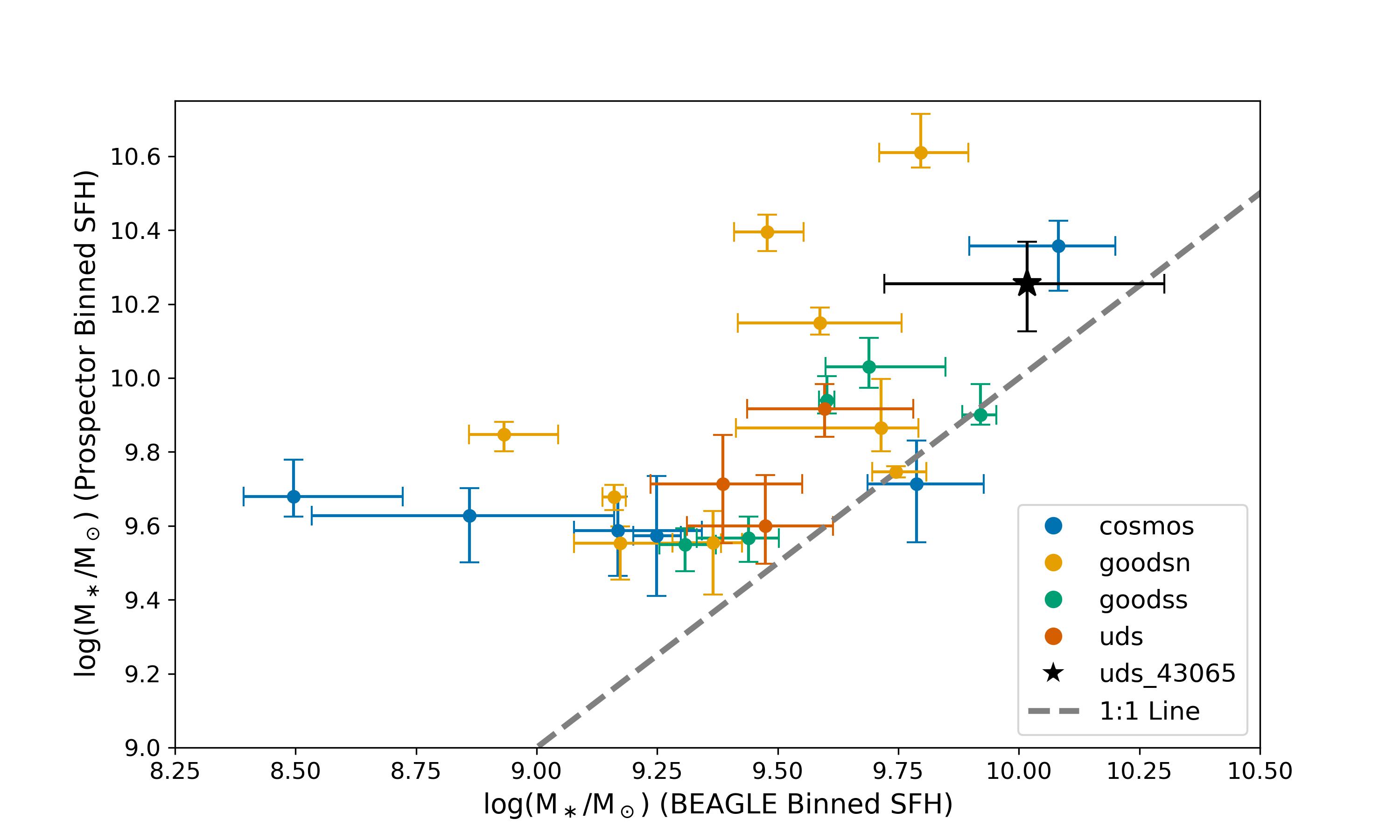} 
    \caption{This plot depicts the \textsc{prospector} and \textsc{beagle} stellar mass estimates of the 24 star-forming Lyman-break $z\sim6$ galaxies in our sample with $M_\ast>3\times 10^9\,M_\odot$ from \textsc{prospector}. We identify two galaxies with $M_\ast >10^{10}\,M_\odot$ in both \textsc{prospector} and \textsc{beagle}, one being the exceptionally red object UDS\_43065, shown here with a black star. } 
    \label{fig:massComparison}
\end{figure*}

The apparent strong Balmer break in UDS\_43065 implies a substantial contribution to the continuum light from relatively old B and A-type stars, while the prominent H$\alpha$ emission implies significant ionizing flux output from younger O stars (see, e.g., \citealt{Endsley2025}).
Crucially, those younger O stars cannot outshine the B and A stars creating the Balmer break.
For this reason, both the \textsc{beagle} and \textsc{propsector} SED fits imply that there is a dearth of early-type O stars (age$<$5 Myr; see Fig. \ref{fig:SFH}) which have immensely high light-to-mass ratios.
The H$\alpha$ photons in these model fits are instead powered by fainter late-type O stars formed in an intense burst of star formation 5--10 Myr ago.
Both \textsc{beagle} and \textsc{prospector} infer that this burst had a very large SFR of $\sim$500–1000 $M_\odot$/yr which produced $\sim$20–40\% of the total stellar mass in UDS\_43065 (see Fig. \ref{fig:SFH}).
The moderate Balmer break produced by the late-type O stars \citep[e.g.,][]{Endsley2025,Trussler2025} is supplemented by that from the dominant population of B and A stars in these models, yielding the red F200W$-$F277W color measured in UDS\_43065.
 
\section{Discussion and Conclusions}\label{sec:discussion}

For over two decades, it has been known that galaxies within the first billion years typically exhibit very blue rest-UV slopes ($-2.5 \lesssim \beta \lesssim -2$; e.g., \citealt{Bouwens2003,Stanway2005,Dunlop2012,Finkelstein2012}). However, with a few rare exceptions \citep[e.g.,][]{Fujimoto2022,Endsley2023_cos87259,Shapley2025}, it has historically been much more challenging to confidently identify and subsequently characterize the properties of very red ($\beta > -1$) objects in the epoch of reionization (see \citealt{Stark2025} for a review). In this work, we used \JWST{}'s greatly improved near-infrared sensitivity to identify and characterize an exceptionally red ($\beta \approx -0.6$) yet UV and optically bright (F115W = 26.0; F444W = 23.9) Lyman-break galaxy at $z_\mathrm{phot} \approx 5.6$ in the UDS/PRIMER field. Here, we discuss how this object fits into the broader context of reionization-era objects, and how the inferred SFH of UDS\_43065 may be key to shedding light on the rapid formation of massive quenched galaxies.

UDS\_43065 is one of only two galaxies with an inferred $M_* > 10^{10}\,M_\odot$ with both \textsc{beagle} and \textsc{prospector} across our full sample of 813 $z\sim6$ LBGs identified over $\approx$500 arcmin$^2$ (see Fig. \ref{fig:massComparison}).
Specifically, we infer \(\log_{10}(M_*/M_\odot) = 10.02^{+0.28}_{-0.30}\) and \(\log_{10}(M_*/M_\odot) = 10.25^{+0.12}_{-0.13}\) with \textsc{beagle} and \textsc{prospector}, respectively. Such a high stellar mass is necessary to explain the observed UV luminosity ($\Muv{} \approx -20.3$) while being so red ($\beta \approx -0.6$).
These stellar mass estimates imply that UDS\_43065 lies in the high-mass exponential tail of the $z\sim6$ galaxy stellar mass function \citep[e.g.,][]{Weibel2024}.

Moreover, UDS\_43065 is the only $z\sim6$ LBG in our sample that confidently shows a rest-frame UV slope of $\beta > -1$ (Fig. \ref{fig:red_galaxy}a).
While several other extremely red dropout objects at $z>5$ have been identified with \HST{} and \JWST{} imaging \citep[e.g.,][]{Fujimoto2022,Greene2024,Matthee2024}, these so-called `Little Red Dots' appear as point-like objects at 4--5$\mu$m and are ubiquitously found to be a sub-class of broad-line AGN.
In contrast, UDS\_43065 is clearly extended in F444W (Fig. \ref{fig:red_galaxy}b; $r_{e} = 400\pm10$ pc) making it exceptional among known extremely red $z\gtrsim6$ Lyman-break objects (see also \citealt{Shapley2025}).

UDS\_43065 is $\approx2-3\times$ smaller than typical $z\sim5-6$ LBGs at similar UV luminosity \citep[e.g.,][]{Shibuya2015,CurtisLake2016}, and is morphologically resolved. 
This relatively small size combined with the exceptionally large stellar mass yields a very high stellar mass surface density ($\Sigma_\mathrm{eff} = M_\ast / (\pi r_e^2)$) of log$(\Sigma_\mathrm{eff}/(M_\odot\,\mathrm{kpc}^{-2}))=10.25\pm0.13$. 
Such a high $\Sigma_\mathrm{eff}$ is comparable to that of massive quiescent galaxies at $z\sim2-3$ \citep{vanDokkum2008} as well as those recently confirmed at $z\sim5-7$ \citep[e.g.,][]{Carnall2023b,deGraaff2025,Weibel2025}. 
At these elevated stellar mass surface densities, feedback processes are expected to weaken, facilitating the rapid conversion of gas into stars \citep[e.g.,][]{Grudic2019}. 
The exceptionally high $\Sigma_\mathrm{eff}$ of UDS\_43065 may therefore offer an important clue as to why it appears to have undergone an intense starburst roughly $\approx$5--10 Myr ago (SFR$\approx$500-1000 $M_\odot$/yr; see Fig. \ref{fig:SFH}). 

Observational campaigns focused on far-infrared wavelengths have long demonstrated that the brightest and most heavily obscured star-forming galaxies at $z>5$ are objects experiencing very dramatic bursts of star formation (SFR$\sim$500--3000 $M_\odot$/yr; e.g., \citealt{Walter2012,Riechers2013,Marrone2018,Casey2019}).
In context of the \textsc{beagle} and \textsc{prospector} SED fits, the very strong dust attenuation within UDS\_43065 ($A_V \approx 1.5$ mag) is manifesting 5--10 Myr after such an intense starburst event (Fig. \ref{fig:SFH}).
But notably, the dust attenuation in UDS\_43065 is relatively moderate enough to keep this object bright in the rest UV and optical, thereby enabling it to be identified via a Lyman-$\alpha$ break selection.
It is possible that the recent $\approx$5 Myr of negligible star formation inferred in UDS\_43065 allowed much of the dust formed in the intense burst to be destroyed or expelled, thereby allowing UDS\_43065 to be selected as a UV-bright LBG, though other physical scenarios cannot be ruled out \citep[e.g.,][]{Esmerian2022,McKinney2025,Narayanan2025}.

If UDS\_43065 indeed experienced a dramatic burst of star formation 5--10 Myr ago, it would represent a transitional phase in which early massive DSFGs evolve into (at least temporarily) passive systems and potentially help resolve the puzzling abundance of early massive quenched galaxies (see \S\ref{sec:intro}). 
The fact that numerous $M_* > 10^{10}\ M_\odot$ quenched $z\gtrsim5$ galaxies are now being identified in pencil-beam \JWST{} fields seems, at face value, challenging to resolve with the fact that bright $z>5$ DSFGs have historically been found in surveys spanning several square degrees (c.f., \citealt{Casey2019, Endsley2023_cos87259}).
But this apparent discrepancy is greatly alleviated if the period of intense star formation within DSFGs is often very short ($\lesssim$5 Myr) and DSFGs frequently transition immediately into a passive phase lasting several tens to hundreds of Myr. 
Because UDS\_43065 was identified over an $\approx$500 arcmin$^2$ ($\approx$0.15 deg$^2$) search area, its discovery may imply that the abundance of massive, currently-passive galaxies that exited a DSFG phase within the past $\approx$10 Myr is far greater than the abundance of active massive DSFGs at $z\gtrsim5$.
Such a scenario would also explain the ubiquitous intense starburst phases inferred from NIRSpec spectra of known $z\gtrsim5$ massive quenched galaxies \citep[e.g.,][]{Carnall2023b,deGraaff2025,Weibel2025}.

Several next steps are required to test this picture properly.
First, NIRSpec follow-up of UDS\_43065 is crucial to validate our inferred recent SFHs from \textsc{beagle} and \textsc{prospector} fits to the deep HSC+ACS+NIRCam photometry.
While we have employed two state-of-the-art SED fitting codes with different underlying SPS models, Bayesian priors, etc., the strong dust attenuation in UDS\_43065 raises the possibility that a significant dust-age degeneracy is impacting our SFH inferences, in which we allow a high degree of flexibility using the non-parametric priors.
Second, ALMA and MIRI follow-up will be necessary to determine if UDS\_43065 harbors optically-obscured star formation activity or an AGN.
Such data will help inform if UDS\_43065 is in fact currently inactive, and if an AGN may have contributed to any recent quenching.
Finally, further dedicated searches for extremely red ($\beta > -1$), bright star-forming galaxies at $z\gtrsim5$ over moderate-to-wide areas ($\gtrsim$0.1 deg$^2$) are necessary to better constrain the space density of systems like UDS\_43065 and how they compare to active, far-IR selected DSFGs.
A promising path forward is pairing deep \textit{Roman} or \textit{Euclid} surveys with overlapping deep multi-band optical imaging from the ground \citep{EuclidCollaboration2025,Weaver2025} while simultaneously exploiting deep long-wavelength far-IR (2--3mm) surveys (e.g., \citealt{Casey2019}).

\section{Acknowledgments}
NAH would like to thank her fellow authors, research advisors, the NSF, and the administrative staff at the University of Texas at Austin for making this research possible, particularly Dr. Ryan Endsley and Prof. John Chisholm for their guidance and feedback throughout the research process. NAH also thanks the FRESCO team for providing additional spectroscopic data that assisted with visual verification of galaxy candidates, as well as Christina Kausek for her feedback. NAH acknowledges support from the NSF REU grant AST-2244278 (PI: Jogee). 

This work is based in part on observations made with the NASA/ESA/CSA James Webb Space Telescope. The data were obtained from the Mikulski Archive for Space Telescopes at the Space Telescope Science Institute, which is operated by the Association of Universities for Research in Astronomy, Inc., under NASA contract NAS 5-03127 for JWST. This research is based in part on observations made with the NASA/ESA Hubble Space Telescope obtained from the Space Telescope Science Institute, which is operated by the Association of Universities for Research in Astronomy, Inc., under NASA contract NAS 5–26555. We thank the numerous HST and JWST survey teams (in particular CANDELS, PRIMER, JADES, and FRESCO) for designing and executing the observing programs that delivered the data used in this work.
We thank the numerous HST and JWST survey teams (in particular CANDELS, PRIMER, JADES, and FRESCO) for designing and executing the observing programs that delivered the data used in this work, and (where applicable) for developing their observing program with a zero-exclusive-access period.
We are grateful for the collective contributions of the approximately 20,000 humans around the world who designed, built, tested, commissioned, and operate JWST.
Some of the data products used herein were retrieved from the Dawn JWST Archive (DJA). DJA is an initiative of the Cosmic Dawn Center (DAWN), which is funded by the Danish National Research Foundation under grant DNRF140.
The authors acknowledge the Texas Advanced Computing Center (TACC) at The University of Texas at Austin for providing HPC resources that have contributed to the research results reported within this paper.

The Hyper Suprime-Cam (HSC) collaboration includes the astronomical communities of Japan and Taiwan, and Princeton University.
The HSC instrumentation and software were developed by the National Astronomical Observatory of Japan (NAOJ), the Kavli Institute
for the Physics and Mathematics of the Universe (Kavli IPMU), the
University of Tokyo, the High Energy Accelerator Research Organization (KEK), the Academia Sinica Institute for Astronomy and
Astrophysics in Taiwan (ASIAA), and Princeton University. Funding was contributed by the FIRST program from the Japanese Cabinet Office, the Ministry of Education, Culture, Sports, Science and
Technology (MEXT), the Japan Society for the Promotion of Science
(JSPS), Japan Science and Technology Agency (JST), the Toray Science Foundation, NAOJ, Kavli IPMU, KEK, ASIAA, and Princeton
University. This paper makes use of software developed for Vera
C. Rubin Observatory. We thank the Rubin Observatory for making their code available as free software at http://pipelines.lsst.io/.
This paper is based on data collected at the Subaru Telescope and
retrieved from the HSC data archive system, which is operated by
the Subaru Telescope and Astronomy Data Center (ADC) at NAOJ.
Data analysis was in part carried out with the cooperation of Center
for Computational Astrophysics (CfCA), NAOJ. We are honored and
grateful for the opportunity of observing the Universe from Maunakea, which has the cultural, historical and natural significance in
Hawaii.

\bibliography{main}{}
\bibliographystyle{aasjournal}

\end{document}